\documentclass[]{aastex}
\usepackage{emulateapj5}

\slugcomment{Accepted for Publication in The Astrophysical Journal}

\shorttitle{Spitzer Photometry of HAT-P-1b}
\shortauthors{Todorov et al.}

\begin{document}


\title{{\it Spitzer} IRAC Secondary Eclipse Photometry of the Transiting Extrasolar Planet HAT-P-1b}


\author{Kamen Todorov\altaffilmark{1,2,3}, Drake~Deming\altaffilmark{2}, 
Jospeph~Harrington\altaffilmark{4}, Kevin~B.~Stevenson\altaffilmark{4},
William~C.~Bowman\altaffilmark{4}, Sarah~Nymeyer\altaffilmark{4}, 
Jonathan~J.~Fortney\altaffilmark{5}, and Gaspar~A.~Bakos\altaffilmark{6,7}}


\altaffiltext{1}{Department of Physics, Astronomy \& Geophysics, Connecticut College,
    Box 5361, \\ New London, CT 06320-4196}
\altaffiltext{2}{Goddard Center for Astrobiology, Goddard Space Flight Center, Greenbelt MD 20771}
\altaffiltext{3}{Present address: Department of Astronomy and Astrophysics,
Pennsylvania State University, University Park, PA 16802}
\altaffiltext{4}{Planetary Sciences Group, Department of Physics, University of Central Florida,\\ 
  4000 Central Florida Blvd., Orlando, FL 32816}
\altaffiltext{5}{Department of Astronomy and Astrophysics, University of California at Santa Cruz, 
 \\ Santa Cruz, CA 95064}
\altaffiltext{6}{Harvard-Smithsonian Center for Astrophysics, Cambridge, MA 02138}
\altaffiltext{7} {NSF Fellow}


\begin{abstract}
We report {\it Spitzer}/IRAC photometry of the transiting giant
exoplanet HAT-P-1b during its secondary eclipse.  This planet lies
near the postulated boundary between the pM and pL-class of hot
Jupiters, and is important as a test of models for temperature
inversions in hot Jupiter atmospheres.  We derive eclipse depths for
HAT-P-1b, in units of the stellar flux, that are:
$0.080\%\pm0.008\%\,[3.6\mu$m], $0.135\%\pm0.022\%\,[4.5\mu$m],
$0.203\%\pm0.031\%\,[5.8\mu$m], and $0.238\%\pm0.040\%\,[8.0\mu$m].
These values are best fit using an atmosphere with a modest
temperature inversion, intermediate between the archetype inverted
atmosphere (HD\,209458b) and a model without an inversion. The
observations also suggest that this planet is radiating a large
fraction of the available stellar irradiance on its dayside, with
little available for redistribution by circulation.  This planet has
sometimes been speculated to be inflated by tidal dissipation, based
on its large radius in discovery observations, and on a non-zero
orbital eccentricity allowed by the radial velocity data. The timing
of the secondary eclipse is very sensitive to orbital eccentricity,
and we find that the central phase of the eclipse is
$0.4999\,\pm0.0005$. The difference between the expected and observed
phase indicates that the orbit is close to circular, with a $3\sigma$
limit of $|e\,\cos\,{\omega}| < 0.002$.

\end{abstract}


\keywords{stars: planetary systems - eclipses - techniques: photometric}

\section{Introduction}

Infrared light from transiting extrasolar planets can be measured
using high precision photometry when the planets pass behind their
stars.  Most of these secondary eclipse measurements to date have used
the {\it Spitzer Space Telescope} \citep{charb05, deming05, deming06,
deming07, demory, harrington07, charb08, machalek, machalek09,
knutson08, knutson09}. {\it Spitzer} data in the four bands of the
IRAC instrument \citep{fazio} define the planet's spectral energy
distribution over the 3.6-8.0\,$\mu$m wavelength region, where water
vapor plays a principal role in shaping the spectrum.  {\it Spitzer}
studies have shown that there are at least two classes of close-in
extrasolar giant planets, differentiated by the temperature gradient
with height in their atmosphere \citep{hubeny, fortney06, fortney,
burrows07, burrows08}.  One class (pM) exhibits a temperature
inversion at high altitude in their atmosphere \citep{knutson08,
knutson09, machalek}.  The inversion affects the emergent spectral
energy distribution by causing the water bands to appear in
emission. The cause of the inversion is believed to be absorption of
strong stellar irradiation in the visible, possibly by TiO/VO
\citep{hubeny, fortney06, fortney}, but other absorbers remain
possible \citep{burrows07, zahnle}.

\citet{machalek} found that XO-1b exhibits an inverted atmospheric
structure, in spite of being nominally below the stellar irradiance
level projected to define the transition between the pM and the
(non-inverted) pL classes. The pM/pL transition corresponds to the
condensation of TiO \citep{fortney}. Recently, \citet{machalek09}
found that XO-2b exhibits a weak temperature inversion, consistent
with being near the pM/pL transition. Like XO-1b and XO-2b, HAT-P-1b
\citep{bakos} lies at the lower edge of that predicted boundary, so it
provides an important additional test of the propensity toward
inversion at that level of irradiation. HAT-P-1b may also exhibit a
non-zero eccentricity to its orbit \citep{johnson}, that may cause
internal energy generation via the dissipation of tidal stress.

In this paper, we report {\it Spitzer} photometry of HAT-P-1b's
secondary eclipse in the four IRAC bands. We use these data to
investigate the atmospheric temperature structure and energy balance
of this planet, and to place more stringent limits on the eccentricity
of its orbit via the timing of the secondary eclipse \citep{charb05}.

\section{Observations}

We observed HAT-P-1b during two secondary eclipses using {\it
Spitzer}/IRAC.  On 2006 December 28, we observed an eclipse at 4.5 and
8.0\,$\mu$m for 356 minutes. To avoid saturation for this relatively
bright star, each frame comprised two 2-second exposures in stellar
mode at 4.5\,$\mu$m (3212 exposures total), simultaneously with one
12-second exposure at 8\,$\mu$m (1606 exposures total). On 2007
December 29, we observed an eclipse at 3.6 and 5.8\,$\mu$m for 333
minutes (1510 total exposures at each wavelength). For that eclipse,
we elected to increase the efficiency at 3.6\,$\mu$m and avoid
saturation by placing the star at the corner of a pixel, thus
spreading the light over 4 pixels.  We used 12-second exposures at
both wavelengths, in full array mode.  For both eclipses, we centered
the star on a portion of the array chosen to avoid known bad pixels
and scattered light from bright stars imaged onto other regions of the
focal plane.

\section{Photometry}

All of our photometry used version S18.5.0 of the Basic Calibrated
Data from the Spitzer pipeline.  We applied the recommended factors to
correct for the variation in flat-field response to point sources
versus extended sources, and the variation in pixel solid angles, as
described in Secs. 5.3 and 5.6.2, respectively, of the IRAC Data
Handbook V3.0.  We ran our analysis both with and without these
corrections, and we find that they change our final results by less
than the $1\sigma$ error in the eclipse depth at all wavelengths.
Their largest effect is at 4.5\,$\mu$m, where they increase the
eclipse depth by $6 \times 10^{-5}$ in units of the stellar flux. All
of the results quoted in this paper include these corrections.

To facilitate the error analysis, we converted the intensities in the
images to electron numbers, using the calibration information in the
FITS headers.  We corrected energetic particle hits by comparing each
pixel to a median-filtered time series of that pixel's intensity,
using a 5-frame resolution. We replaced individual values exceeding
the median by more than $4\sigma$ with the median value.  Some
energetic particle hits that overlie the stellar image were not well
corrected, and resulted in outlying intensities in the photometric
time series; those values were omitted when fitting eclipse curves to
the data.

All of our results are based on aperture photometry, but the details
differ with wavelength, as described below.  At all wavelengths, we
centered the aperture on each stellar image by fitting a parabola to
the brightest three points in the stellar profile.  The image was
summed in X to define the profile as a function of Y, and vice versa.
We also found the center of each image by fitting a 2-dimensional
Gaussian to the core of the image point spread function, but this did
not improve the results over the parabolic fit.  We varied the size of
the photometry aperture, and used the value that minimized the scatter
in the time series.  

At each wavelength, we subtracted the background due to solar system
zodiacal thermal emission. Since HAT-P-1 has a brighter companion star
11 arcsec distant, there was also a contribution due to diffracted
light from the companion. Based on model PRFs provided by the Spitzer
Science Center, we expect a diffracted contribution of $\sim
0.3$MJy/sr at the position of HAT-P-1.  This contribution is only
weakly dependent on wavelength, because the decreasing brightness of
the star at longer wavelength works in the opposite direction to the
effect of diffraction.  However, the zodiacal background is
significantly wavelength dependent, so diffracted light from the
companion star can dominate at 3.6 and 4.5\,$\mu$m, decreasing to
$\sim 15$\% of the zodiacal background at 8\,$\mu$m.  We therefore
measured the background at a symmetric position on the opposite side
of the companion star from HAT-P-1, using an aperture of the same size
and shape.  Since this relatively small aperture (typically about 3
pixels in radius) encompasses relatively few electrons from the
background, we increase the precision of the background measurement by
fitting a parabola to the background time series, and using the value
of that fit for each frame.  Apart from random error, the background
variation was quite gradual and was well represented by the parabolic
fit.

\subsection{3.6 and 4.5\,$\mu$m}

{\it Spitzer} photometry at 3.6 and 4.5\,$\mu$m is known to be
affected by pixel position, wherein the value from aperture photometry
is a function of the location of the stellar centroid within the pixel
\citep{morales, charb05}. While this is true for our photometry at
these wavelengths, an additional factor is important for our
3.6\,$\mu$m observations.  The upper panel of Figure~1 shows the
photometric intensity of the star at 3.6\,$\mu$m, as a function of the
Y-position (the X-position effects were not significant). The bulk of
the data were collected with the star displaced significantly from
pixel center, as was our intention when planning the observations.
The median value for the maximum per-pixel intensity in these images
is about 130,000 electrons, approximately equal to the 1\%
non-linearity point for this detector (see the Spitzer Observing
Manual, Sec. 6.1.3).  However, due to pointing jitter the star
sometimes wanders closer to pixel center, and the intensity drops as
saturation begins to affect the data (red points on Figure 1).  Those
points were not included in our analysis.  The remaining data still
exhibit variations in intensity that are dependent on pixel position,
but the variation is complex because of the location of the star near
pixel boundaries. We required a sixth-order polynominal in Y-position
to fit this relation; incorporating the X-position in the fit did not
improve it. We subtracted the fit (blue line on Figure 1) from the
unsaturated photometric points, to decorrelate the variation due to
pixel position.

The lower panel of Figure~1 plots the uncorrected intensity versus
orbital phase, showing that the eclipse is visible even without the
correction for pixel position. This panel also shows that the omitted
points (in red) occur periodically as the oscillation in telescope
pointing moves the stellar centroid toward the center of the pixel.
Since most of the correlation between intensity and pixel position
depends on the Y-coordinate of the star, we used a square photometric
aperture at both 3.6 and 4.5\,$\mu$m, including fractional pixels at
the edge. We reasoned that a square aperture would provide the
cleanest isolation of pixel-position effects that are predominately
dependent on the Y-coordinate. We found that the photometric scatter
at 3.6\,$\mu$m was minimized using an aperture 5 pixels on a side (2.5
pixels in `radius').

Our photometry at 4.5\,$\mu$m shows a much weaker pixel position
effect (not illustrated) than at 3.6\,$\mu$m, but does not involve
saturation.  The 4.5\,$\mu$m data show a normal dependence on distance
from pixel center \citep{morales}, but again with a stronger dependence on
the Y- than the X-coordinate.  The correlation coefficient between the
image $\delta Y$-position and intensity is -0.3, indicating a weak
correlation, but statistically significant considering the 1604
degrees of freedom.  We removed this correlation with a linear
relation between intensity and radial distance, and the minimum
scatter was achieved using a box 7 pixels on a side (3.5 pixels
`radius').  We explored other methods to achieve this decorrelation,
such as using a function of Y only, and both X and Y separately, but
the minimum $\chi^2$ was achieved using radial distance.  Figure~2
shows the 3.6 and 4.5\,$\mu$m photometry plotted as a function of
orbital phase, with the decorrelation functions overplotted.  Note that
the decorrelation functions are smooth when plotted versus pixel
position, but they become more jagged when plotted versus orbital phase
as in Figure~2.

Both the 3.6 and 4.5\,$\mu$m photometry achieve a precision near the
photon limit after the decorrelations.  Specifically, the scatter at
3.6\,$\mu$m is $0.00131$, which is merely 6\% greater than the
photon-limited value.  At 4.5\,$\mu$m, the scatter of $0.00533$
exceeds the photon-limit by 11\%.  The eclipse fitting procedure at
every wavelength always used the unbinned data, but some plots show
binned data for clarity. Figure~3 shows binned photometry for both 3.6
and 4.5\,$\mu$m, with the best-fit eclipse curves and $\pm 1\sigma$
error limits.

\subsection{5.8 and 8.0\,$\mu$m}

Photometry at 5.8 and 8.0\,$\mu$m did not exhibit a detectable pixel
position effect, there being no significant correlation between intensity
and either the X- or Y-coordinate of the image.  We performed the
photometry using both square and circular apertures.  Although our
results do not depend significantly on the adopted shape of the
photometric aperture, we elected to use a circular aperture at both
5.8 and 8.0\,$\mu$m.  We found a minimum scatter in the photometry
using aperture radii of 2.4 and 2.8 pixels at 5.8 and 8.0\,$\mu$m,
respectively.

\subsection{Eclipse Amplitudes}

Following the background subtraction and aperture photometry, we
divide the time series at each wavelength by its average value.  This
places the results in contrast units, i.e., relative to the stellar
flux.

We generated an eclipse curve numerically, using the stellar and
planetary parameters from \citet{winn} and \citet{johnson}.  The
numerical code was tested for transit curves by comparing to the
analytic formulae given by \citet{mandel}, and it attains an accuracy
of $10^{-6}$, more than sufficient for our purpose (an eclipse curve
has the same shape as a transit curve in the limit of zero
limb darkening).  In fitting to each set of photometry, we scale the
depth of the eclipse curve and vary its central phase, but leave its
shape (duration, limb-crossing time) unchanged. At all wavelengths, we
fit a baseline curve plus the eclipse curve simultaneously via
multi-variable linear regression, but the nature of the baseline curve
varies with wavelength (see below).  Since the linear regressions
cannot fit a variable central phase, we perform the regressions
separately for each of a series of central phase values, and we pick
the solution having the minimum $\chi^{2}$.

We found that a linear baseline was adequate for the eclipse fits at
all wavelengths except 8.0\,$\mu$m where the well-known detector ramp
\citep{charb05, deming06, harrington07, knutson08, knutson09, desert}
exhibits a quasi-logarithmic shape.  We model the 8\,$\mu$m ramp,
$R(\phi)$, as a sum of a linear and logarithmic term in phase
($\phi$):

\begin{equation}
R(\phi)=a_{0}+a_{1}\phi+a_{2}ln(\phi-\phi_{0})
\end{equation}

We adopt multiple values for $\phi_{0}$, and solve for $a_{0}$,
$a_{1}$, and $a_{2}$ by linear regression at each adopted value for
$\phi_{0}$.  Note that $\phi_0$ is a phase offset used to facilitate
the ramp fit, and is not related to the orbit of the planet. We choose
the best fit from the 2-D grid of $\phi_{0}$ and eclipse central phase
values based on the minimum $\chi^{2}$.  We found that this model of
the ramp provides consistently excellent fits, but the best-fit
$a_{i}$ and $\phi_0$ can be degenerate in the sense that different
combinations can produce indistinguishable ramps.  Fortunately, we did
not find the eclipse depth and central phase to exhibit significant
degeneracies with the ramp parameters. Figure~4 shows the 5.8 and
8\,$\mu$m data before ramp removals, with the best-fit ramps and
eclipses overlaid. Figure~5 shows the eclipse fits at 5.8 and
8\,$\mu$m, in comparison to binned data with the baseline and ramp
effects removed.

Note that a possible temporal drift in intensity at 3.6 and
4.5\,$\mu$m is a phenomenon physically distinct from the pixel
position effect.  After correcting intensity for pixel position, we
include a linear baseline when fitting the eclipse curves. We have
added these linear baselines to the total decorrelation function
illustrated in Figure~2. The slope of the baseline at 3.6\,$\mu$m is
$0.018\%\pm0.003\%$ per hour. \citet{knutson09} found a linear increase
at 3.6\,$\mu$m of similar magnitude. We found that all of the
intensity variations at 4.5\,$\mu$m were fully accounted for by
changes in pixel position, and the temporal term was not significant.

Best-fit eclipse depths and errors, in units of the stellar flux
(contrast units) are given in Table~1.

\subsection{Error Estimation}

We estimate the errors using the bootstrap Monte Carlo technique
\citep{press}.  The bootstrap technique generates synthetic data sets
using the residuals from the best-fit model, and permutes them to
make new data.  Each new boostrap data set is constructed as follows.
To construct N new data points, we start with the $i=1,N$ points from
the best-fit curve.  We draw a residual randomly from the pool of
original residuals, add it to the $i^{th}$ best-fit curve point,
return that residual to the pool, and draw again until we have created
a data set of $N$ new points.  We make 50,000 data sets using this
procedure.

For each new bootstrap data set, we repeated the entire fitting
procedure, with the exception of the pixel position fits.  We did this for
the 50,000 bootstrap data sets at each wavelength, and tabulated
histograms of the eclipse depth, baseline parameters, and central
phase. These histograms are very close to Gaussian distributions, and
their standard deviations give estimates of the error in the fit
parameters.

We also estimated errors using the residual permutation method
described by \citet{southworth}.  This method - sometimes called the
`prayer bead' method - is similar to the bootstrap technique, except
that it preserves the order of the residuals, and is therefore more
sensitive to the presence of red noise. In most cases the error from
the permutation method was very close to the bootstrap error.  The
adopted error for a given parameter was taken to be the greater of the
values from the bootstrap and residual permutation method, and Table~1
includes these errors for the eclipse depth and central phase.

\section{Results and Discussion}

\subsection{Atmospheric Temperature Structure}

Figure~6 shows our contrast values plotted versus wavelength, and also
the results from \citet{knutson08} for HD\,209458b, the archetype of
an inverted atmosphere.  We include the contrast predicted by
two 1-D models of the HAT-P-1b planetary atmosphere \citep{fortney05,
fortney}, both invoking re-emission of stellar irradiance on the dayside
only. The solid line is the nominal model, without a temperature
inversion. This nominal model is self-consistent and non-gray, with
solar metallicity, and was calculated as in \citet{fortney}. It has
a bond albedo of 0.067 and a dayside effective temperature of
1512K. This model has no inversion because we find that it is too cool
to allow gas-phase TiO.

The dashed line is a weakly inverted model, also of solar metallicity.
It is {\it not} self-consistent; it has a constant, ad-hoc temperature
inversion of $dT/d\log P = -30$K.  However, this model is constrained
to have the same effective temperature (1512K) as the non-inverted
model.  Since the temperature gradient of the inverted model is
shallow, the emission features are weak, and it falls close to the
contrast expected from a 1500K blackbody (dot-dashed line).  In all
instances, we used the planetary and stellar radii from \citet{winn},
and a Kurucz 6000/4.5/0.0 model to represent the stellar spectrum
\citep{torres}.

The rotation of HAT-P-1b should be tidally locked to its orbit, even
though its orbital period is longer than for many planets in the hot
Jupiter class.  Using Eq.(1) of \citet{guillot}, we calculate a
spin-down time of $8 \times 10^6$ years, starting with Jupiter's
rotation rate and adopting a very conservative Q-value ($10^6$).
Since this spin-down time is much less than the age of the system
\citep{torres}, we expect tidal locking of the planet's rotation.

Based on tidally-locked rotation, the maximum dayside temperature of
HAT-P-1b is 1550K, assuming zero albedo, a uniform temperature over
the dayside hemisphere, and no transport to the nightside. This is too
cool to produce an inversion using TiO/VO absorption, so we first
investigate the non-inverted model.  Emission at the three longest
IRAC wavelengths can arise from levels higher in the atmosphere than does
the 3.6\,$\mu$m radiation \citep{burrows07}, and these three channels
exhibit a contrast that is moderately higher than the non-inverted
model.  Comparing to the HD\,209458b results \citep{knutson08} shows
good agreement at 3.6 and 8.0\,$\mu$m, but HAT-P-1b exhibits a
contrast at 4.5 and 5.8\,$\mu$m that is intermediate between
HD\,209458b and the non-inverted model.  This seems qualitatively
consistent with a moderate inversion, perhaps produced by an absorber
other than TiO/VO.

We integrated the flux from each planetary atmospheric model and the
stellar model atmosphere over the IRAC bandpass functions.  Dividing
these integrated fluxes at each wavelength produces expectation values
for the observations. (These are not illustrated on Figure~6, but they
fall very close to the contrast values at the band-center
wavelengths.) We used these expectation values to calculate the
$\chi^2$ value for the observations compared to each model.  For the
inverted model (dashed line), $\chi^2$ is $3.9$, whereas it is $11.0$
for the non-inverted model.  A value as high as $11.0$ will occur only
$2.6$\% of the time if the non-inverted model is an accurate
description of HAT-P-1b's atmosphere.  This level of confidence is not
sufficient to rigorously eliminate the non-inverted model, but it does
indicate that the inverted model is preferable.

Within the errors, our results for HAT-P-1b can also be reproduced using a
blackbody spectrum for the planet (dot-dashed line on Figure~6, $\chi^2 =3.5$). 

The luminosity of HAT-P-1A is 1.48 times solar \citep{torres}, and it
receives a stellar flux $\sim 2/3$ of the HD\,209458b case.  If
HAT-P-1b absorbs with zero albedo and re-emits uniformly but only on
the dayside hemisphere, then we expect a maximum temperature of 1550K.
If the planet's emission is uniform over both hemispheres, we expect
an observed temperature of 1300K. Our observations at secondary
eclipse are best described by a blackbody having a temperature of
$1500\pm100$K, where we have factored in the random error of our
observations as well as uncertainty in the stellar and planetary
parameters. However, most of the flux is probably emitted at shorter
wavelengths \citep{barman, seager}, not directly observable using {\it
Spitzer}.  At face value our results suggest that redistribution of
the stellar irradiation by dynamics may be inefficient for this
planet.
				  
\subsection{Orbital Eccentricity}

HAT-P-1b orbits a star in a wide visual binary, and this circumstance
can have significant consequences for the orbital dynamics of the
planet. The inclination and eccentricity ($e$) of the planet's orbit can in
principle undergo oscillations and long term evolution due to the
Kozai mechanism \citep{fabrycky}.  Misalignments between the planet's
orbital inclination and the rotation axis of the star can result, and
can be observed using the Rossiter-McLaughlin effect. \citet{johnson}
find that the angle between the sky projections of stellar spin axis
and the orbit normal for HAT-P-1b is $3.7\pm2.1$ degrees.  They also put an upper
limit on the eccentricity of $0.067$ with 99\% confidence.  These
values suggest that any Kozai oscillations of HAT-P-1b have largely
damped out. Our results for the secondary eclipse timing will further
strengthen that conclusion, as discussed below.

The timing of the secondary eclipse is exquisitely sensitive to the
eccentricity of the orbit, with one ambiguity being the value of
$\omega$, the argument of periastron. When our line of sight aligns
with the minor axis of the planet's orbit ($\omega =0$ or $\pi$), then
the secondary eclipse will not be centered on phase $0.5$ unless the
orbit is circular.  When our line of sight aligns with the major
axis of the orbit ($\omega = {\pi}/2$ or $3{\pi}/2$), departures from
circularity will affect the duration of the eclipse, but not the
central phase.  The phase of secondary eclipse therefore constrains
the value of $e\,{\cos}\,\omega$ \citep{charb05}.

We observed two eclipses, each at two wavelengths simultaneously.  The
difference in best fit phases for the same eclipse observed at different
wavelengths is approximately consistent with our errors (Table~1).
The larger difference occurs at 3.6 and 5.8\,$\mu$m, where the central
phase difference is $0.5016-0.4992=0.0027$.  Because the noise at each
wavelength is independent, the error on the difference in the two
phases is the quadrature sum ($=0.0014$) of the phase errors at the
two wavelengths. Hence the phase difference is less than a $2\sigma$
variation.

Weighting the central phase of the eclipse at each wavelength
(Table~1) by the inverse of its variance, we find a mean value of
$0.4999\pm0.0005$.  Considering the 55 seconds of light travel time
across the orbit, we expect to find the eclipse at phase $0.500014$.
Accounting for the light travel time and the error in the observed
eclipse phase, we derive a $3\sigma$ upper limit of
$|e\,\cos\,{\omega}| < 0.002$.  If this planet's orbital eccentricity
was affected by Kozai oscillations in the past, they have damped to
the point where the orbit is closely circular.

\acknowledgements This work is based on observations made with the
{\it Spitzer Space Telescope}, which is operated by the Jet Propulsion
Laboratory, California Institute of Technology, under a contract with
NASA.  Support for this work was provided by NASA.  We are grateful to
the anonymous referee for thoughtful comments that improved this
paper.



{\it Facilities:} \facility{Spitzer}.




\clearpage



\begin{figure}
\epsscale{.45}
\plotone{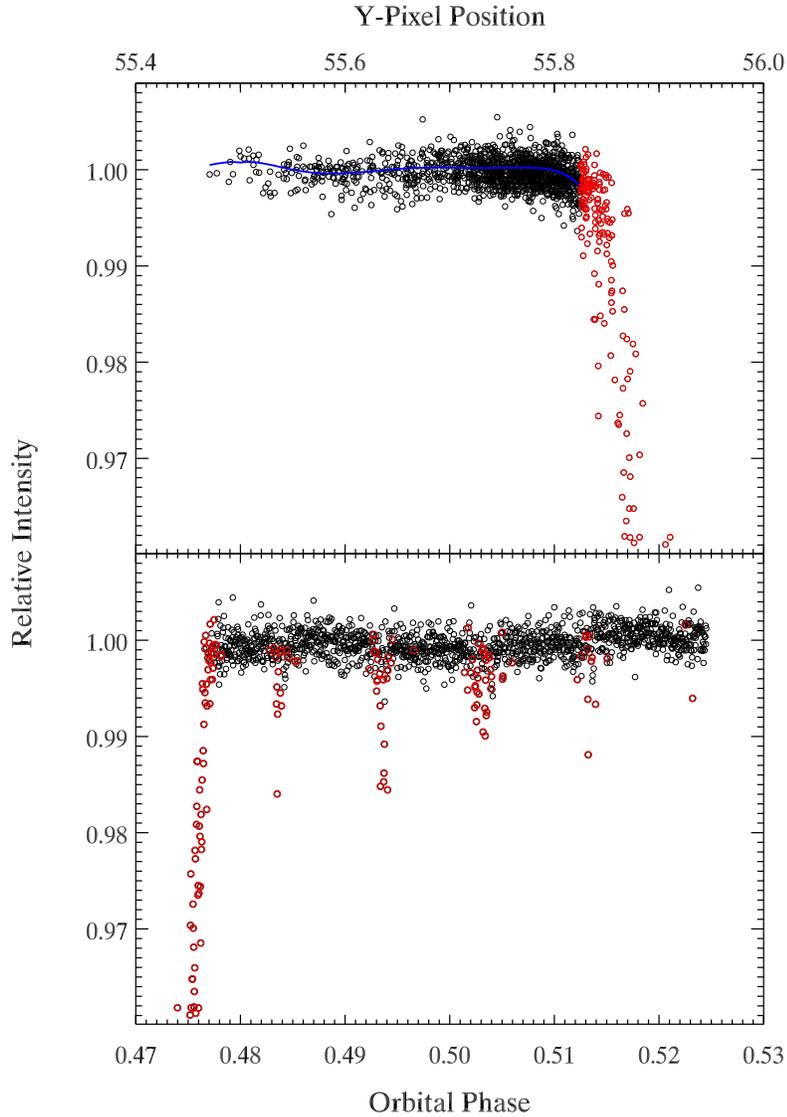}
\vspace{0.7in}
\caption{{\it Upper panel:} Photometric intensity versus Y-pixel
  position for 3.6\,$\mu$m photometry, before correction. Pixel
  centers are at integer coordinate values.  The red points represent
  lower intensities due to approaching saturation for this bright star
  as it moves closer to pixel center in the Y coordinate, and those
  points were not used in our analysis. The blue line is the fitted
  correction function.  The X-position of the star (not illustrated)
  was approximately pixel 79. (Both X- and Y-pixel coordinates are
  1-based.) {\it Lower panel:} Intensity versus orbital phase for the
  3.6 $\mu$m photometry, before the correction. The red points are not
  used in the analysis. Note that the eclipse is visible in this plot,
  but see Figs.~2~and~3 for a clearer view.
  \label{fig1}}
\end{figure}

\clearpage

\begin{figure}
\epsscale{.50}
\plotone{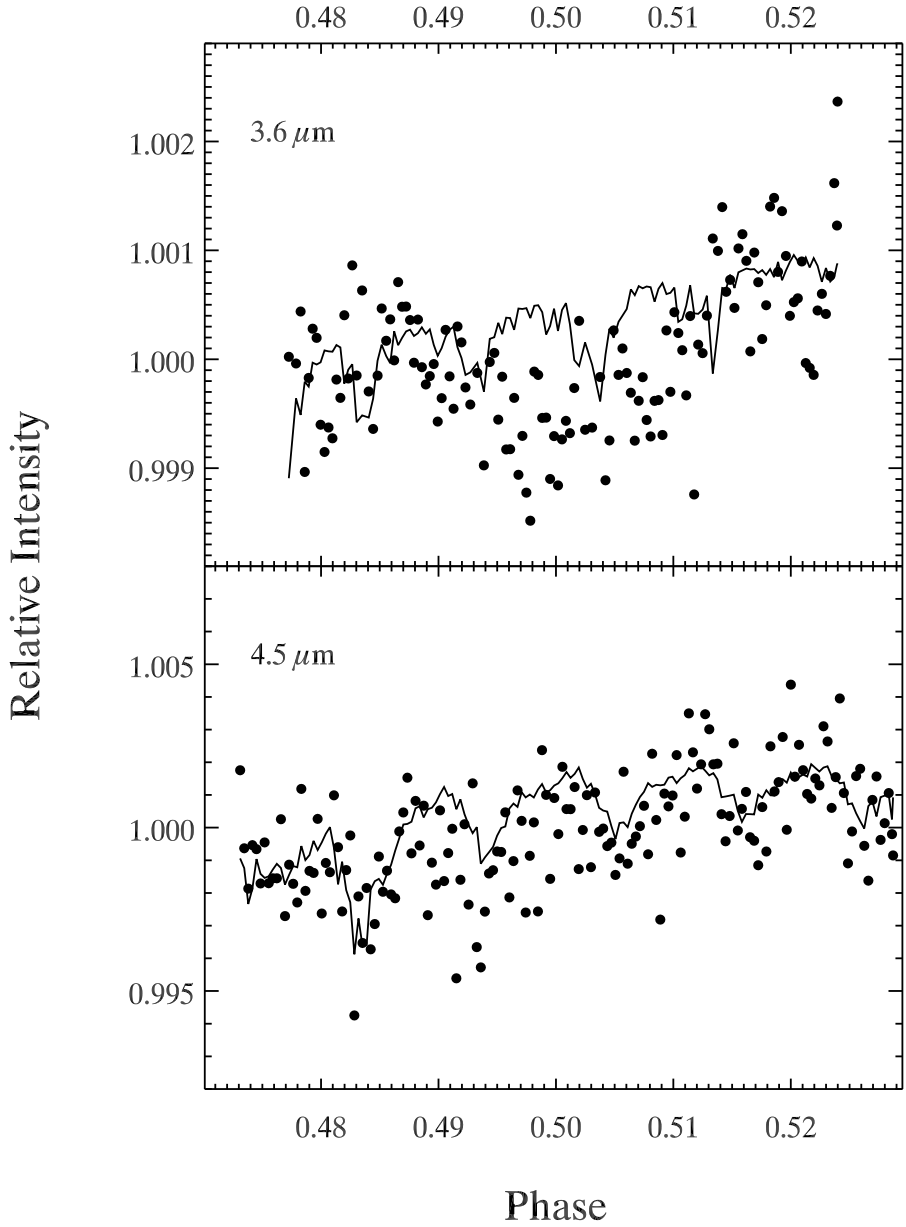}
\vspace{0.8in}
\caption{Photometry at 3.6\,$\mu$m (upper panel) and 4.5\,$\mu$m
 (lower panel), before decorrelation with image position.  For
 clarity, each plotted point represents the binned average of the
 original photometry.  In the upper panel each bin represents 10
 exposures (about 2.4 minutes per bin). In the lower panel, each bin
 represents 20 exposures (about 2.2 minutes per bin).  The lines show
 the decorrelation functions.  Note that these functions are smooth
 when plotted as a function of pixel phase, but because of pointing
 jitter they exhibit fluctuations when plotted here as a function of
 orbital phase.  The eclipses are clearly visible near orbital phase
 0.5, as differences between the points and lines.  The decorrelation
 function at 3.6\,$\mu$m includes a linear drift as a function of time
 (see text), but all of the variations at 4.5\,$\mu$m can be
 attributed to changes in pixel position.
\label{fig2}}
\end{figure}

\clearpage

\begin{figure}
\epsscale{.65}
\plotone{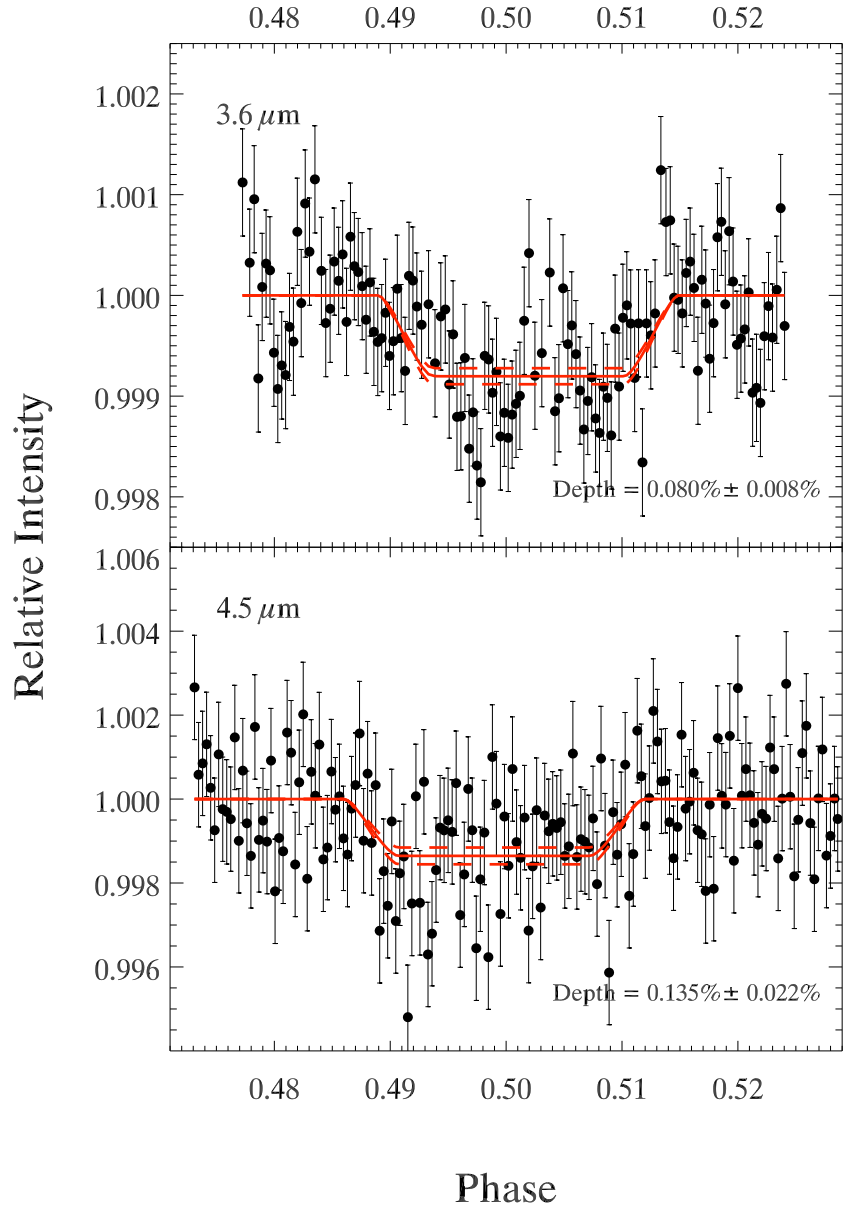}
\vspace{0.5in}
\caption{Binned photometry at 3.6\,$\mu$m (upper panel, 10 exposures
  per bin) and 4.5\,$\mu$m (lower panel, 20 exposures per bin),
  corrected for the pixel phase effect (see Figures~1 \& 2, and
  text). The solid red line shows the fitted secondary eclipse, and
  the dashed red lines show the $\pm\ 1\sigma$ range on the eclipse
  depth, from the bootstrap Monte-Carlo trials (see Table~1, and
  text). The error bars give the standard deviation of the mean, 
  based on the internal scatter within each bin.
\label{fig3}}
\end{figure}

\clearpage

\begin{figure}
\epsscale{.50}
\plotone{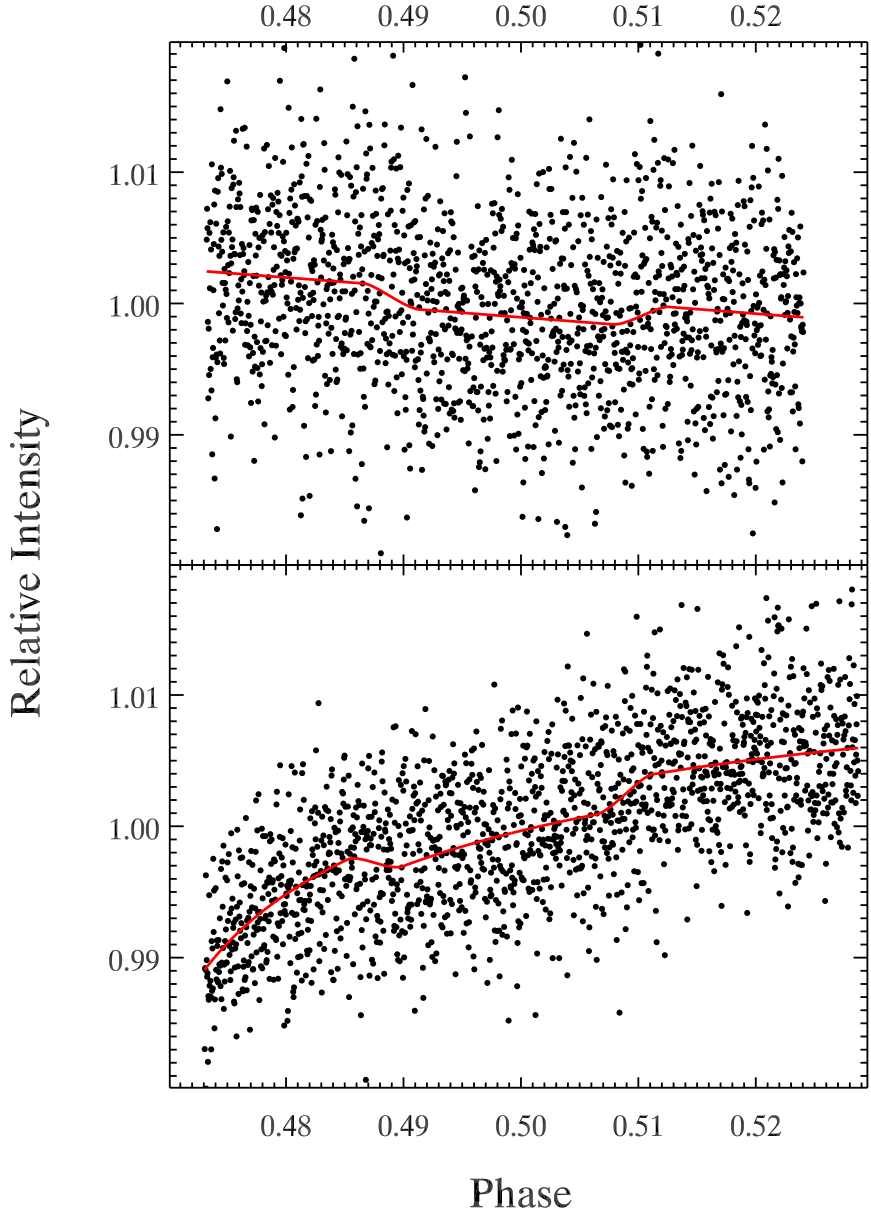}
\vspace{0.8in}
\caption{Photometry at 5.8\,(upper panel) and 8.0\,$\mu$m (lower
panel) before removal of the detector ramp. The points are
unbinned. The solid red lines are the best fit baseline ramps plus
eclipse curves, obtained via linear regression (see text). A linear
ramp was used at 5.8\,$\mu$m, whereas at 8.0\,$\mu$m the ramp is
comprised of a linear plus logarithmic term, and the best-fit (see
text) is $1.005 \times 10^{-2}ln(\phi-0.465)-0.07\phi+1.07063$, where
$\phi$ is phase.
\label{fig4}}
\end{figure}

\clearpage

\begin{figure}
\epsscale{.45}
\plotone{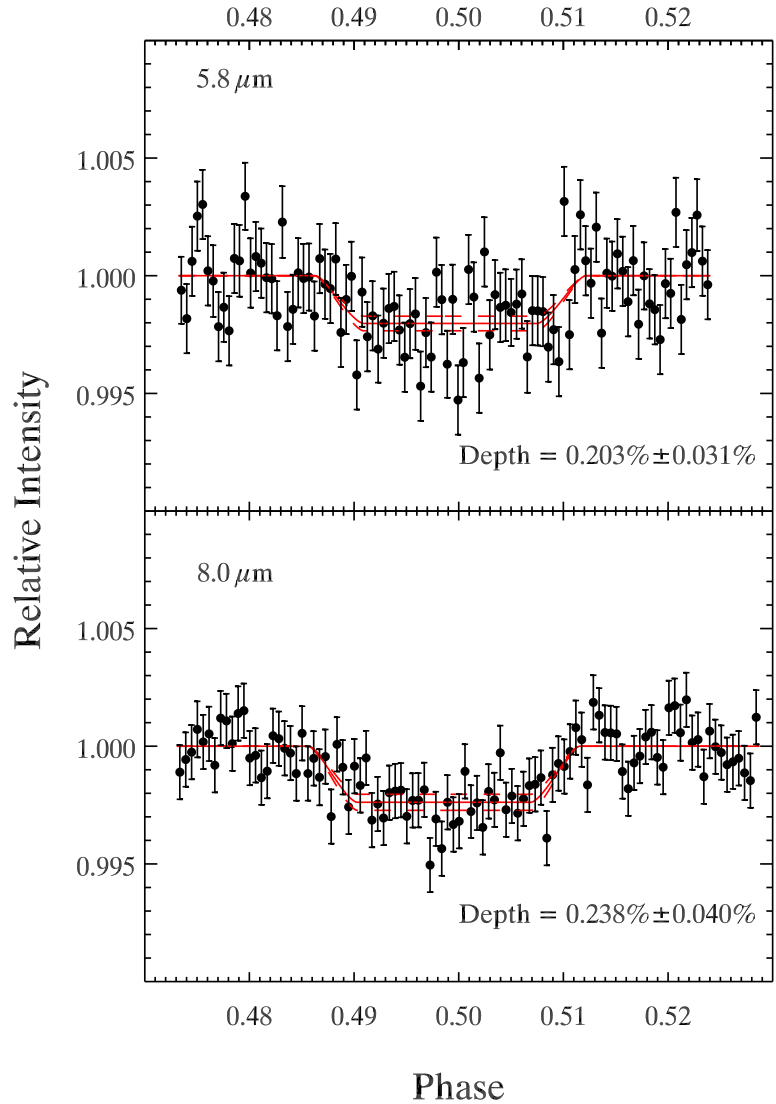}
\vspace{0.8in}
\caption{Binned photometry at 5.8\,$\mu$m (upper panel) and 8\,$\mu$m
(lower panel), with the baselines removed.  The bin size is 16 points
at both wavelengths, about 3.6 minutes per bin, but the eclipse fit
used the unbinned data (see Figure~4). The solid red lines show the
fitted secondary eclipse, and the dashed red lines show the $\pm\
1\sigma$ range on the eclipse depth, from the bootstrap Monte-Carlo
trials (see Table~1, and text). The error bars give the standard deviation 
of the mean, based on the internal scatter within each bin.
\label{fig5}}
\end{figure}

\clearpage

\begin{figure}
\epsscale{.80}
\plotone{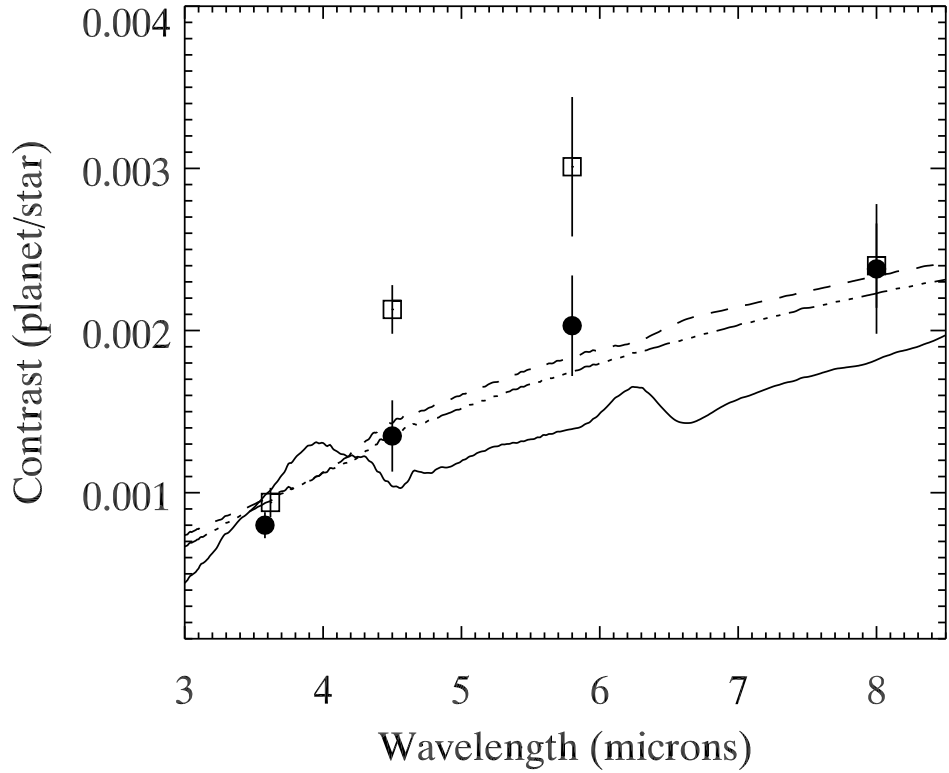}
\vspace{0.5in}
\caption{Contrast values for HAT-P-1b (solid circles) versus
wavelength. The open squares are for HD\,209458b, from
\citet{knutson08}. The two observed points at 3.6\,$\mu$m have been
offset slightly in wavelength to prevent overlap of their error bars.  The solid
line is a model for the planet \citep{fortney} with no temperature
inversion, and with re-distribution of stellar irradiance over the
dayside only - with none to the nightside. The dashed line is for a
model with a modest temperature inversion (see text). In the inverted
model, very modest decreases in contrast can be seen at the
wavelengths of contrast peaks in the non-inverted model. The
dot-dashed line is the contrast for a 1500K blackbody having the same
radius as HAT-P-1b \citep{winn, torres}.
\label{fig6}}
\end{figure}

\clearpage

\begin{table}
\begin{center}
\caption{Fitted Eclipse Depth, Central Phase and Heliocentric Julian Date
(HJD) of Mid-eclipse. \label{tbl-1}}
\begin{tabular}{crrc}
\tableline\tableline
Wavelength & Eclipse Depth & Central Phase & HJD-2454000\\
\tableline
 3.6 $\mu$m & $0.080\%\pm0.008\%$ & $0.5016\pm0.0008$ & $464.4228\pm0.0036$ \\
 4.5 $\mu$m & $0.135\%\pm0.022\%$ & $0.4991\pm0.0010$ & $102.7229\pm0.0045$ \\
 5.8 $\mu$m & $0.203\%\pm0.031\%$ & $0.4992\pm0.0012$ & $464.4121\pm0.0054$ \\
 8.0 $\mu$m & $0.238\%\pm0.040\%$ & $0.4986\pm0.0010$ & $102.7206\pm0.0038$ \\
\tableline
\end{tabular}
\end{center}
\end{table}



\clearpage




\end{document}